\documentclass[preprint,prl,showpacs]{revtex4}
\usepackage{graphicx}
\usepackage{dcolumn}
\usepackage{bm}

\begin{document}
\preprint{APS/123-QED}
\author{R. K\"uchler$^{(1)}$, P. Gegenwart$^{(1)}$, K.
Heuser$^{(2)}$, E.-W.~Scheidt$^{(2)}$, G.R. Stewart $^{(3)}$, and
F. Steglich$^{(1)}$}
\address{$^{(1)}$ Max-Planck Institute for Chemical Physics
of Solids, D-01187 Dresden, Germany
\\ $^{(2)}$EP III ,Institut f\"ur Physik, Universit\"atsstrasse 1, 86159 Augsburg, Germany
\\ $^{(3)}$University of Florida, Gainesville, FL 32611-8440, USA}
\title{Gr\"uneisen Ratio Divergence at the Quantum Critical
Point in CeCu$_{6-x}$Ag$_x$}

\date{
\today%
}

\begin{abstract}
The heavy fermion system CeCu$_{6-x}$Ag$_x$ is studied at its
antiferromagnetic quantum critical point, $x_c=0.2$, by low
temperature ($T\geq 50$ mK) specific heat, $C(T)$, and volume
thermal expansion, $\beta(T)$, measurements. Whereas $C/T\propto
\log(T_0/T)$ would be compatible with the predictions of the
itinerant spin-density-wave (SDW) theory for two-dimensional
critical spin fluctuations, $\beta(T)/T$ and the Gr\"uneisen
ratio, $\Gamma(T)\propto \beta/C$, diverge much weaker than
expected, in strong contrast to this model. Both $C$ and $\beta$,
plotted as a function of the reduced temperature $t=T/T_0$ with
$T_0=4.6$ K are similar to what was observed for
YbRh$_2$(Si$_{0.95}$Ge$_{0.05}$)$_2$ ($T_0=23.3$ K) indicating a
striking discrepancy to the SDW prediction in both systems.
\end{abstract}

 \maketitle

The nature of the quantum critical point (QCP) at which long-range
magnetic order gradually develops in heavy-fermion (HF) systems
has been the focus of research activities during past years
\cite{Stewart}. In paramagnetic systems like CeCu$_6$, a Fermi
surface formed by heavy quasiparticles has been detected,
indicating an itinerant $4f$-electron state \cite{Reinders}. The
central question is how these heavy quasiparticles evolve if
these materials are tuned into a long-range magnetically ordered
state. In the traditional picture \cite{Hertz,Millis,Moriya}, the
quasiparticles retain their itinerant character and form a
spin-density wave (SDW) type of antiferromagnetic (AF) ordering.
For this Hertz-Millis-Moriya theory, in the following refered as
{\it itinerant} scenario, a mean-field type of quantum critical
behavior arises. Recent experiments have shown that at least in
some HF systems this picture fails \cite{Schroeder,Custers}.
Consequently, a new type of QCP has been proposed at which the
quasiparticles break-up into their components: conduction
electrons and local $4f$-moments forming magnetic order. This {\it
locally-critical} scenario arises due to the destruction of the
Kondo resonance at the QCP \cite{Si,Coleman}.

In CeCu$_6$, the weakening of the $4f$-conduction electron
hybridization, $J$, induced by a lattice expansion in
CeCu$_{6-x}$M$_x$ with M=Au ($x_c=0.1$) \cite{Loehneysen}, Ag
($x_c=0.2$) \cite{Fraunberger,Heuser98a}, Pd ($x_c=0.05$) and Pt
($x_c=0.1$) \cite{Sieck} drives the system at a critical
concentration $x_c$ towards a QCP beyond which long-range AF
order appears. Remarkably, the non-Fermi liquid (NFL) behavior
observed at the respective $x_c$ in the different systems does
not depend on which dopand M has been used \cite{Sieck}. In fact,
the specific heat coefficient of all of these different systems
is very similar (within $\sim 10\%$ below 3 K) and shows a
characteristic $C/T \propto \log(T_0/T)$ dependence over nearly
two decades in $T$ down to 50 mK. Thus, a common origin of the
quantum critical state at $x=x_c$ for the different systems
CeCu$_{6-x}$M$_x$ is likely. The logarithmic divergence of the
quasiparticle mass would only be compatible with the itinerant
scenario if strongly anisotropic, two-dimensional (2D) critical
spin fluctuations would lead to singular scattering at the whole
Fermi surface \cite{Rosch}. Indeed, inelastic neutron scattering
experiments on CeCu$_{5.9}$Au$_{0.1}$, revealed rod-like
structures of high intensity in $q$-space translating to quasi-2D
fluctuations in real space \cite{Stockert}. The 2D SDW picture,
however, cannot explain that both at and far away from the wave
vector of the nearby AF order the neutron and magnetization data
exhibit $E/T$ and $B/T$ scaling with an anomalous exponent
$\alpha<1$ \cite{Schroeder}. The momentum independence in the
critical response led to the proposal of local criticality
\cite{Schroeder,Si,Coleman}. So far it is not clear, which kind of
mass divergence would occur in such a scenario. At the AF QCP in
YbRh$_2$(Si$_{0.95}$Ge$_{0.05}$)$_2$ a stronger than logarithmic
mass divergence has been observed at very low temperatures,
incompatible with the itinerant scenario \cite{Custers}.
Furthermore, local moments have been detected in the
low-frequency bulk susceptibility \cite{Custers}, as well as in
microscopic magnetic measurements \cite{Ishida,Sichelschmidt},
indicating local criticality in this system. Comparing the title
Ce system with this Yb system, two important questions arise: i)
is there a similar discrepancy in thermodynamic properties with
respect to the predictions of the itinerant scenario for the Ce
system as found \cite{Custers} for
YbRh$_2$(Si$_{0.95}$Ge$_{0.05}$)$_2$ and ii) do both materials
show similar mass divergences? This Letter addresses these
questions.

Very recently, it has been shown that the thermal expansion,
$\beta=V^{-1}(dV/dT)$ ($V$: sample volume), compared to the
specific heat, behavior is much more singular in the approach to
the QCP.
Complementary to the specific heat, being related to the
temperature dependence of the entropy, the thermal expansion
probes the pressure dependence of the entropy and thus should
become singular in the approach of a (pressure-driven) QCP.
Scaling analysis \cite{Zhu} revealed that the Gr\"uneisen ratio
$\Gamma \propto \beta/C$ can be used as a highly sensitive probe
of quantum criticality because it has to diverge at any QCP. The
critical exponent $\epsilon$ in the divergence $\Gamma\propto
1/T^\epsilon$ is given by $\epsilon=1/\nu z$ with $\nu$, the
critical exponent for the correlation length, $\xi\propto
|r|^\nu$ ($r$: distance from the QCP) and $z$, the dynamical
critical exponent in the divergence of the correlation time,
$\tau_c\propto\xi^z$ \cite{Zhu}. For a 3D AF QCP \cite{2D
Grueneisen} the itinerant scenario predicts $\nu=1/2$ and $z=2$
yielding $\epsilon=1$. Thus a study of the Gr\"uneisen ratio can
serve as a basis for a detailed assessment of the validity of the
SDW picture in different systems.

The first-ever observation of a Gr\"uneisen ratio divergence has
been made on the two HF systems CeNi$_2$Ge$_2$ and
YbRh$_2$(Si$_{0.95}$Ge$_{0.05}$)$_2$, both located at AF QCPs
\cite{Kuechler}. For the former system $\epsilon=1$ has been found
within 50 mK $\leq T \leq 3$ K in accordance with the itinerant
scenario. For YbRh$_2$(Si$_{0.95}$Ge$_{0.05}$)$_2$, on the other
hand, the measured Gr\"uneisen exponent is fractional:
$\epsilon=0.7 \pm 0.1$ for $T \lesssim 0.8$ K \cite{Kuechler}.
This cannot be explained by the itinerant theory, but is
qualitatively consistent with the locally quantum critical
picture. As this scenario has first been proposed on the basis of
the neutron scattering experiments on CeCu$_{6-x}$M$_x$ (M=Au)
\cite{Schroeder}, it would be highly desirable, to determine the
critical exponent of the Gr\"uneisen ratio divergence in this
system at the critical concentration $x_c$.

Below, we show first that the QCP in CeCu$_{6-x}$Ag$_x$
($x_c=0.2$) is phenomenologically related to that in
CeCu$_{6-x}$Au$_x$ ($x_c=0.1$) and afterwards investigate the
nature of this QCP by means of a Gr\"uneisen analysis.
Polycrystalline samples of CeCu$_{6-x}$Ag$_x$ with $0.09\leq x
\leq 1.2$ were prepared via arc melting together stoichiometric
amounts of high-purity elements under purified argon atmosphere
\cite{HeuserDok,Heuser98a,Scheidt}. They were analyzed by X-ray
powder diffraction and found to be single phase and with the
proper orthorhombic structure. In contrast to the Au system where
for $x\leq 1$ only one of the five different Cu sites is occupied
by the dopand, the Ag atoms are randomly distributed on all Cu
sites \cite{HeuserDok}. The linear enhancement of the $b$- and
$c$-lattice parameters is observed to be of similar size as in the
Au system. On the other hand, at $x=1$ the $a$-parameter
enhancement is 4\% larger compared to that in the Au system.
Specific heat and thermal expansion measurements were performed in
dilution refrigerators with the aid of a thermal relaxation
calorimeter \cite{Heuser98a} and capacitive dilatometer
\cite{Kuechler}, respectively.

Figure 1 displays the low-temperature specific heat divided by
temperature, $C(T)/T$, of various CeCu$_{6-x}$Ag$_x$ samples on a
logarithmic temperature scale. Long-range AF order is observed
for $x\geq 0.3$ and is manifested by broadened jumps in $C(T)/T$.
The inset shows $T_N(x)$ as determined by (entropy-conserving)
equal areas constructions and the maximum of the derivative
$d\rho(T)/dT$ in corresponding electrical resistivity measurements
\cite{HeuserDok}. Extrapolation of $T_N$ to zero temperature
yields a critical concentration $x_c=0.2$. This value is in
accordance with the extrapolation of the entropy at the AF phase
transition, $S(T_N)$, measuring the size of the ordered moment,
towards zero. Thus, the critical concentration is twice as large
as in the case of the CeCu$_{6-x}$Au$_x$ system although the
relative volume expansion by Ag doping is slightly larger as in
the Au case. It is unlikely that the larger value of $x_c$ is
caused by stronger site disorder in the Ag system. Another
possibility is that the Ag ions adopt a different valence state
than the Au ions \cite{HeuserDok}. As the volume of Ag$^{1+}$ ions
is larger than that of Au$^{2+}$ ions this would explain the
relatively larger volume expansion for Ag, compared to Au.
Furthermore, a change in the electronic configuration might
increase the density of states at the Fermi level. This would
effectively enhance $J$ and stabilize the paramagnetic ground
state leading to a higher $x_c$ for the Ag system
\cite{HeuserDok}.

For the CeCu$_{5.8}$Ag$_{0.2}$ sample which is located right at
the QCP, the specific heat coefficient is very similar to that in
CeCu$_{5.9}$Au$_{0.1}$ and, at zero magnetic field, diverges
logarithmically between 50 mK and 2.5 K (Figure 2a). The
application of magnetic fields leads to the recovery of Landau
Fermi-liquid behavior, $C/T=\gamma(B)$, at lowest temperatures as
observed in many other NFL systems \cite{Stewart}. We observe a
logarithmic divergence $\gamma(B)\propto-\log B$ (not shown)
which represents a clear signature of a QCP at $B=0$. The
zero-field temperature dependence, as well as the field
dependence of $\gamma(B)$, assuming a linear relation between the
magnetic field and the control parameter, would be compatible
with the itinerant theory for an AF QCP, if one assumes truely 2D
critical spin fluctuations.

We now turn to the thermal expansion, measured along three
perpendicular orientations on the same polycrystal studied by
specific heat (Figure 2b). The observed anisotropy is caused by
the texture of the polycrystalline sample. The volume expansion
coefficient $\beta$ is given by the sum of the three linear
expansion coefficients $\alpha_i$ all showing a similar
temperature dependence. Upon cooling to the lowest temperatures,
$\beta(T)/T$ increases strongly and diverges logarithmically for
$T\leq 0.8$ K. Although the observed $\beta/T$ divergence is
steeper than in $C/T$, indicating that the NFL behavior is caused
by a QCP, it is much weaker than the temperature dependence
expected in the 2D-SDW scenario,
$\beta(T)/T=a_0+a_1/T+a_2/T\log\log(T_0/T)$ \cite{Zhu}.

Figure 3 displays the temperature dependence of the Gr\"uneisen
ratio $\Gamma(T)$ calculated from the specific heat and thermal
expansion data shown in Figure 2. In the entire temperature
range, the divergence is weaker than $\Gamma\propto 1/T$ and thus
incompatible with the predictions of the itinerant scenario for
both 3D or 2D critical spin fluctuations \cite{Zhu}. As shown in
the inset, $\Gamma(T)$ roughly follows a logarithmic increase
upon cooling from below 1 K.

In the last part of this paper, we compare the NFL behavior in
CeCu$_{5.8}$Ag$_{0.2}$ with that observed in
YbRh$_2$(Si$_{0.95}$Ge$_{0.05}$)$_2$ \cite{Custers,Kuechler}. As
discussed by Sereni {\it et al.} \cite{Sereni} for a substantial
number of HF systems located near the magnetic instability the
logarithmic increase of the specific heat coefficient $C/T$
involves a similar amount of entropy (about 55\% of $R\ln2$). On
a reduced temperature scale $t=T/T_0$, where the spin-fluctuation
temperature $T_0$ is assumed to be close to the Kondo temperature
$T_K$ determined, e.g. from the residual quasielastic neutron
line width, the specific heat coefficient is described by
$C/t=-D\log t+ET_0$ with $D=7.2$ Jmol$^{-1}$K$^{-1}$
\cite{Sereni}. Here the parameter $E$ is small or negligible for
systems showing a low Kondo temperature as well as excited
crystal electric field states that are well separated from the
ground state. In Figure 4, we show that our specific heat data
agree perfectly with the scaling function using $T_0=4.6$ K (see
straight line). Similar scaling behavior has been reported for
the CeCu$_{6-x}$Au$_x$ series as well \cite{Sereni}. The specific
heat data of YbRh$_2$(Si$_{0.95}$Ge$_{0.05}$)$_2$ collapse for
$t\geq 0.01$ on the scaling function with $T_0=23.3$ K (see
Figure 4). Indeed, this value agrees well with the single-ion
Kondo temperature $T_K\approx 25$ K for this system
\cite{Custers}. For $t<0.01$ the specific heat coefficient of
YbRh$_2$(Si$_{0.95}$Ge$_{0.05}$)$_2$ shows a pronounced upturn
that is not a precursor of the tiny AF order at $T_N=20$ mK
\cite{Custers}. Above 50~mK, the zero-field data of both specific
heat and thermal expansion are identical to their counterparts at
the critical field $B_c=0.027$~T \cite{Kuechler}. Since the
stronger than logarithmic mass divergence in this compound is at
strong variance to the predictions of the itinerant model, it
would be highly desirable to probe the specific heat of the
Ce-based systems at $t\leq0.01$. For CeCu$_{5.8}$Ag$_{0.2}$, this
corresponds to the temperature range below 50 mK which has,
unfortunately, not been investigated so far.

Finally, we turn to the scaling behavior observed in the thermal
expansion. In contrast to the specific heat, the thermal expansion
of CeCu$_{5.8}$Ag$_{0.2}$ deviates strongly from the predictions
of the itinerant theory for $T\geq 50$ mK. In Figure 5 we
demonstrate that, using the same $T_0$ values obtained from the
specific heat scaling, the thermal expansion shows scaling
behavior for both systems as well. As mentioned before, the SDW
theory in the 2D case would predict a very strong divergence
$\beta/T \propto T^{-1}\log\log T$. By contrast, a logarithmic
increase is observed upon cooling over more than one decade in
$t$ for $0.015\lesssim t \lesssim 0.25$, followed by a $1/t$
divergence in the Yb system. Already the weak logarithmic
divergence {\it above} $0.015t$ rules out the 2D SDW scenario for
the QCP in both systems.

In summary, the thermodynamic behavior in CeCu$_{5.8}$Ag$_{0.2}$,
which is located right at an AF QCP, has been studied by
measurements of the specific heat and the thermal expansion. The
observed logarithmic temperature dependence of $C/T$ would be
compatible with the itinerant theory assuming the presence of 2D
critical spinfluctuations. Such a scenario, can however clearly be
excluded from the analysis of the thermal expansion and
Gr\"uneisen ratio data. Both properties diverge towards zero
temperature much weaker than predicted by the itinerant theory.
On the reduced temperature scale $T/T_0$, with $T_0\approx T_K$,
the Kondo temperature, one finds universal behavior in the
thermodynamic properties of both CeCu$_{5.8}$Ag$_{0.2}$ and
YbRh$_2$(Si$_{0.95}$Ge$_{0.05}$)$_2$ for which latter material a
locally critical QCP had been highlighted recently
\cite{Custers,Sichelschmidt,Kuechler}.

We are grateful to C. Geibel, J. Sereni and Q. Si for stimulating
discussions. Part of the work at Dresden was supported by the
Fonds der Chemischen Industrie (Dresden). Work at Augsburg was
supported by the Deutsche Forschungsgemeinschaft (SFB 484).

\newpage
\begin{figure}
\centerline{\includegraphics[width=1\textwidth]{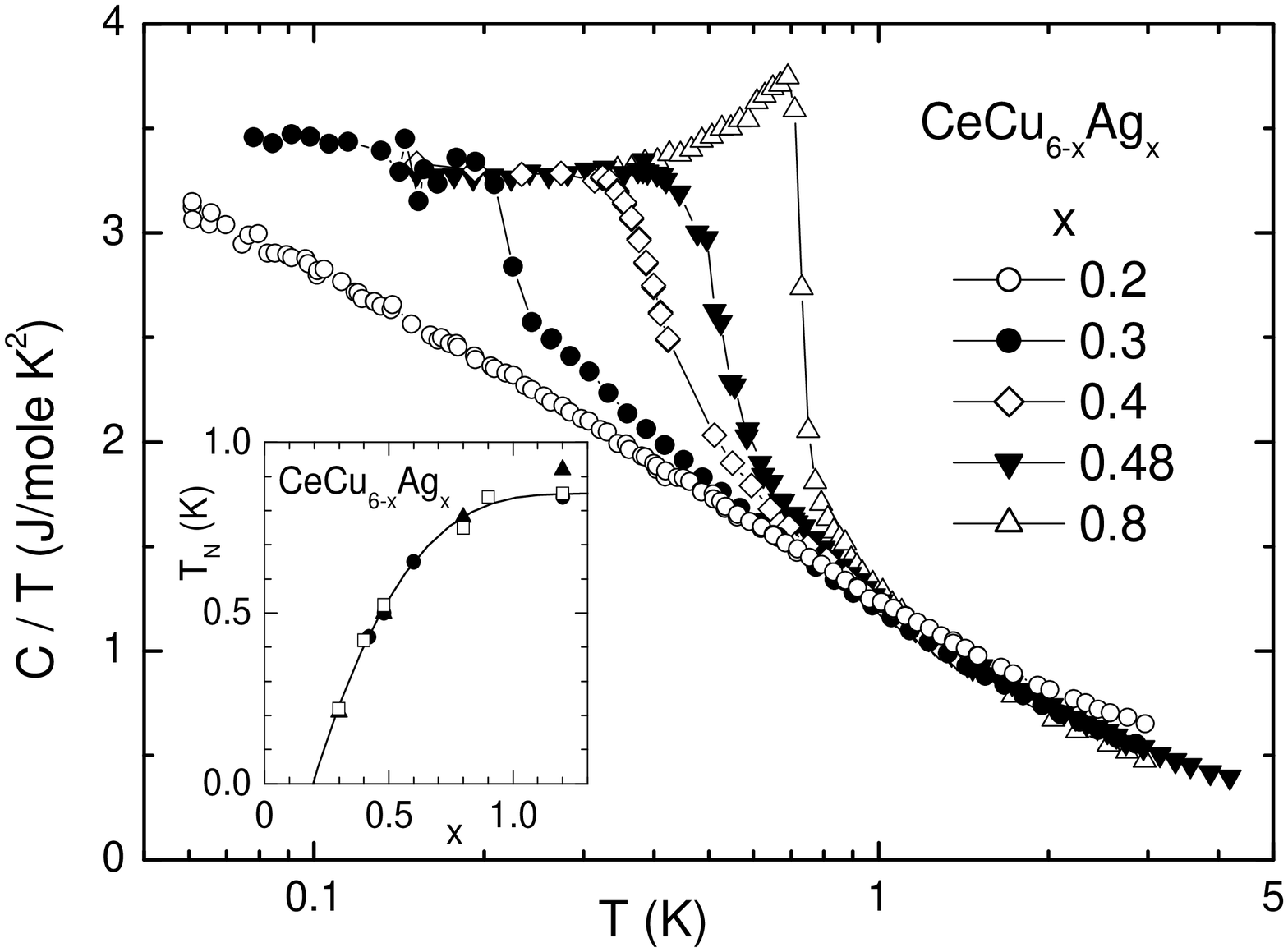}}
\caption{Specific heat as $C/T$ vs $T$ (on a logarithmic scale)
for different CeCu$_{6-x}$Ag$_x$ polycrystals. Inset shows
evolution of antiferromagnetic phase transition temperature $T_N$
vs $x$ as derived from specific heat (squares: this study,
circles \cite{Fraunberger}) and electrical resistivity (triangles
\cite{HeuserDok}) results.} \label{fig1}
\end{figure}

\begin{figure}
\centerline{\includegraphics[width=1\textwidth]{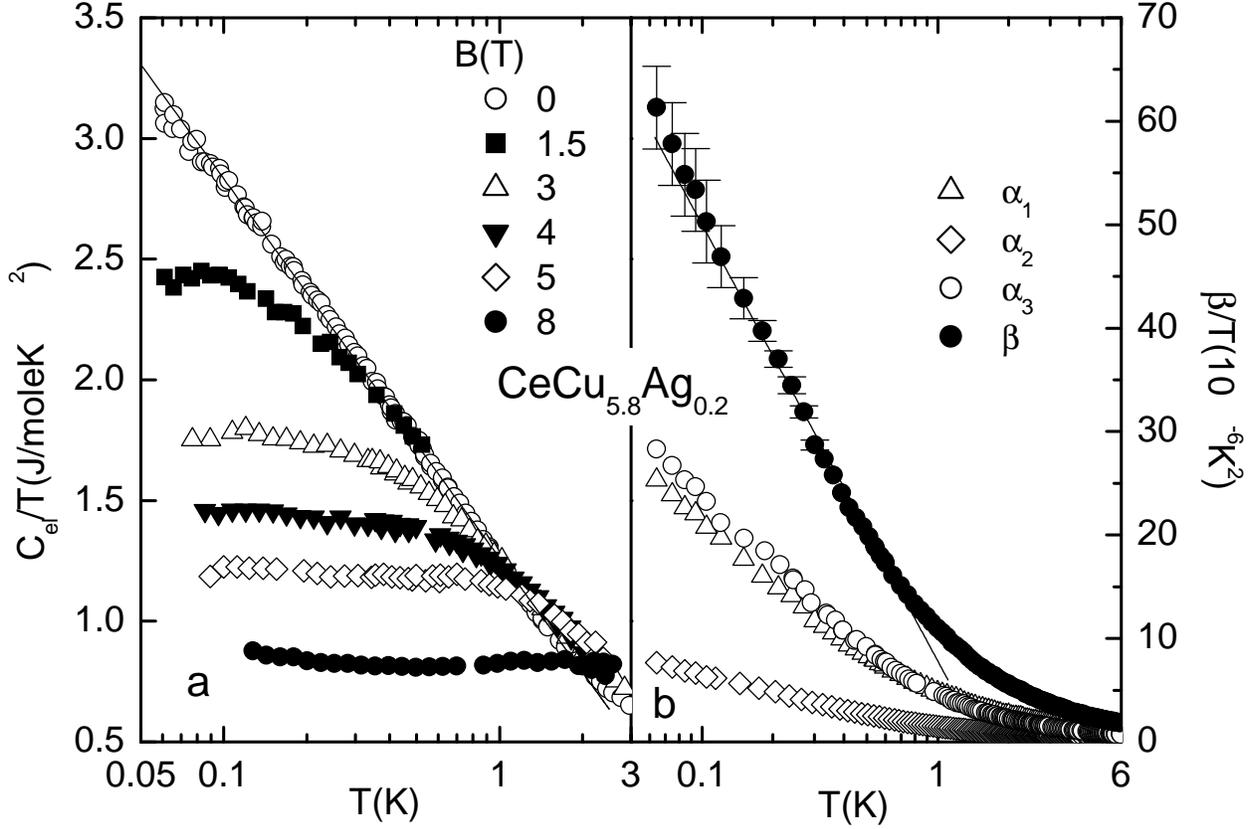}}
\caption{(a) Electronic specific heat of CeCu$_{5.8}$Ag$_{0.2}$
as $C_{el}/T$  vs $T$ (on a logarithmic scale) for $B=0$ and
differing magnetic fields. At $B>0$, $C_{el}$ is obtained after
subtraction of the Cu nuclear specific heat contribution $C_n
\propto B^2/T^2$ \cite{HeuserDok}. (b) Volume thermal expansion
coefficient $\beta$ of the same sample studied in (a) as
$\beta/T$ vs $\log T$. $\beta=\alpha_1+\alpha_2+\alpha_3$ with
$\alpha_i$ being the linear thermal expansion coefficients along
the three perpendicular directions of the sample. Solid lines
indicate logarithmic temperature dependences in $C_{el}(T)/T$ and
$\beta(T)/T$.} \label{fig2}
\end{figure}

\begin{figure}
\centerline{\includegraphics[width=1\textwidth]{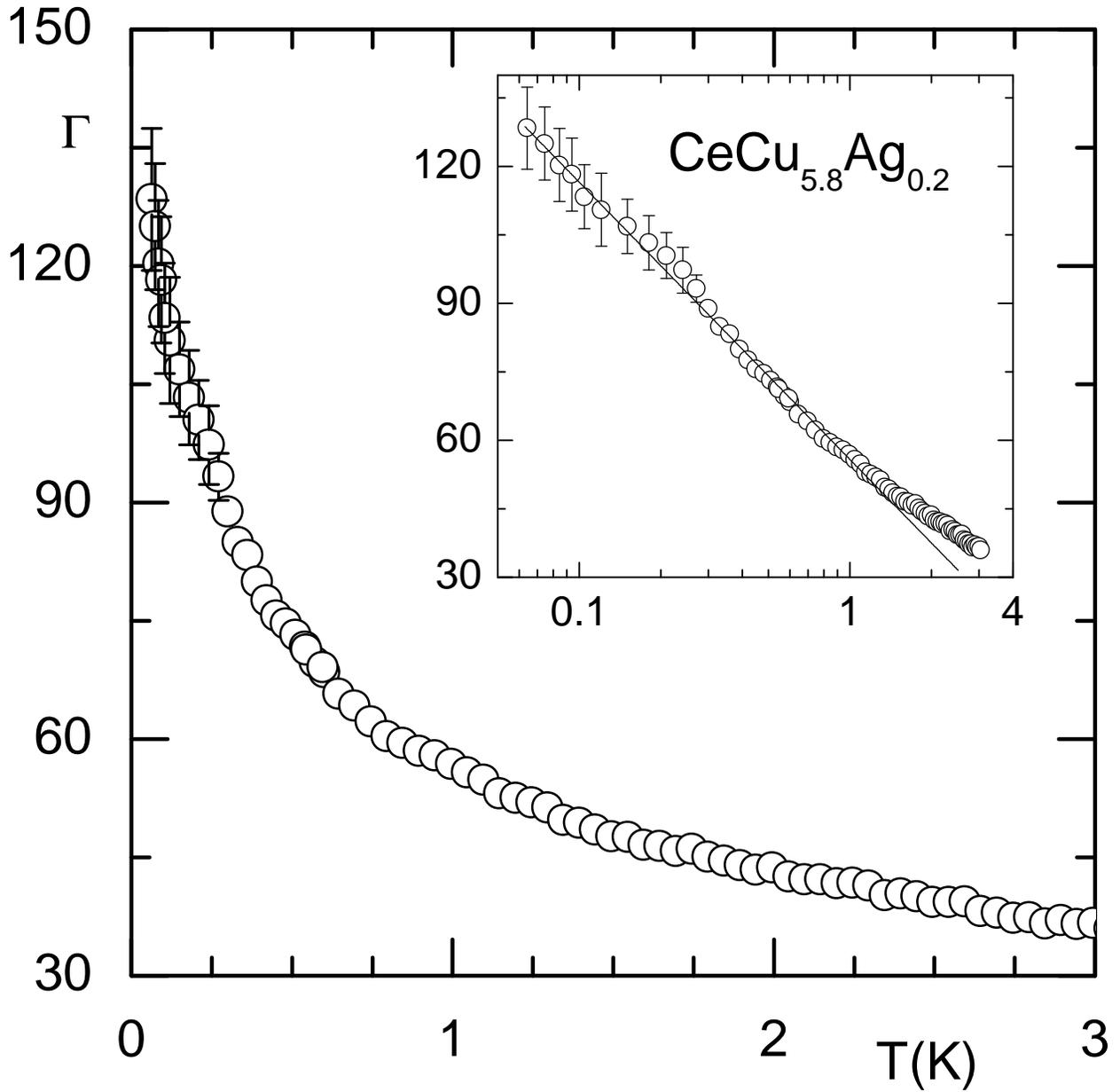}}
\caption{Temperature dependence of Gr\"uneisen-ratio
$\Gamma=V_{mol}/\kappa_T\cdot\beta/C$ with molar volume
$V_{mol}=6.37\cdot 10^{-5}$ m$^3$mol$^{-1}$ and isothermal
compressibility $\kappa_T=1\cdot 10^{-11}$ Pa$^{-1}$ \cite{Oomi}.
Inset shows same data on a logarithmic temperature scale. Solid
line represents $-\log(T)$ dependence.} \label{fig3}
\end{figure}

\begin{figure}
\centerline{\includegraphics[width=1\textwidth]{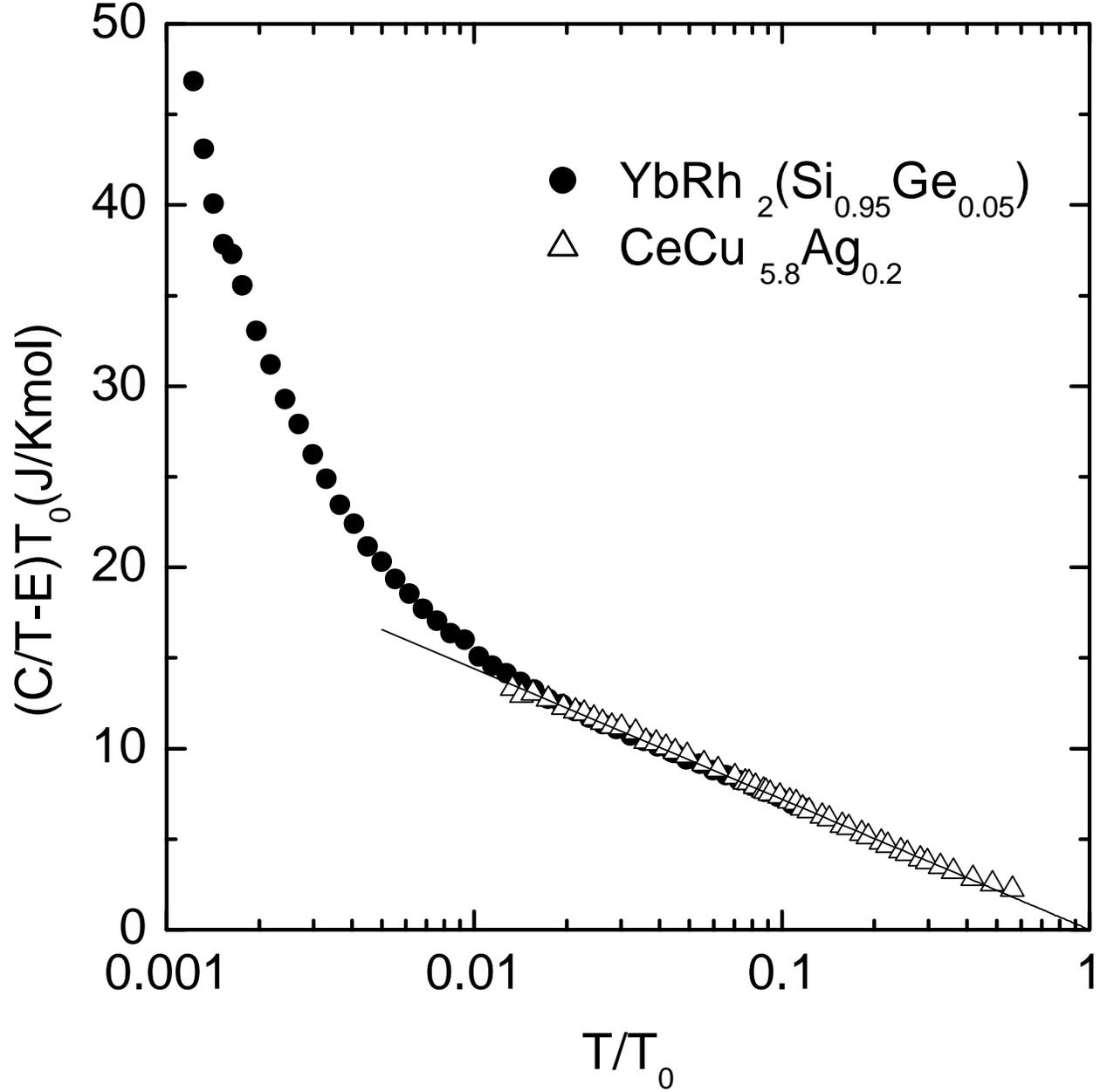}}
\caption{Scaling of the low-$T$ specific heat of
CeCu$_{5.8}$Ag$_{0.2}$ and YbRh$_2$(Si$_{0.95}$Ge$_{0.05}$)$_2$
\cite{Custers,Kuechler} as $(C/T-E)T_0$ vs $\log(T/T_0)$
according to Sereni {\it et al.} \cite{Sereni}. For
CeCu$_{5.8}$Ag$_{0.2}$, $T_0=4.6$ K and $E=0.105$ J/K$^2$mol, for
YbRh$_2$(Si$_{0.95}$Ge$_{0.05}$)$_2$, $T_0=23.3$ K and $E=0.066$
J/K$^2$mol. Solid line represents scaling function
$7.2\log(T_0/T)$ observed in several other HF systems
\cite{Sereni}.} \label{fig4}
\end{figure}

\begin{figure}
\centerline{\includegraphics[width=1\textwidth]{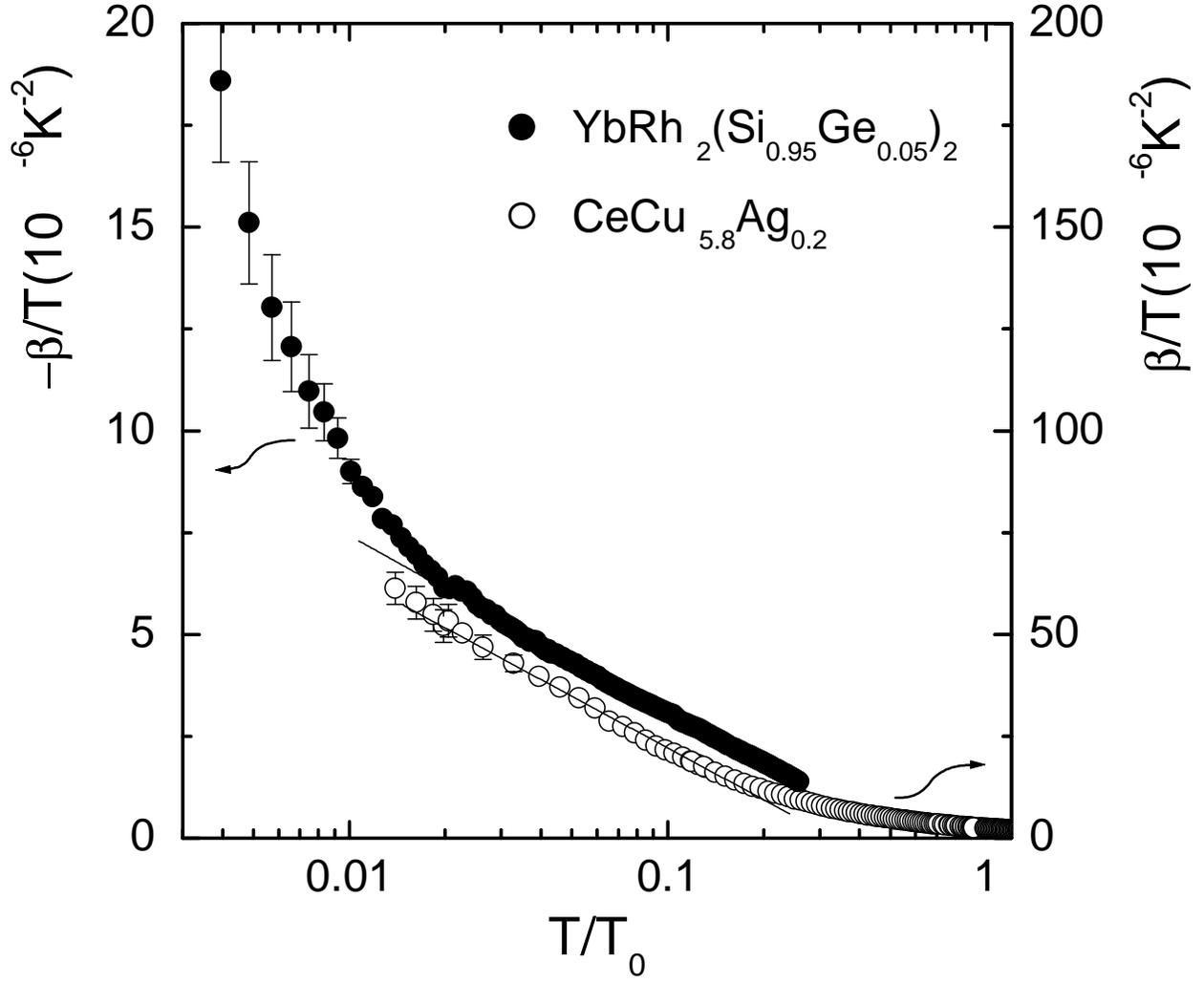}}
\caption{Volume thermal expansion as $\beta/T$ vs normalized
temperature $T/T_0$ (on a logarithmic scale) for
CeCu$_{5.8}$Ag$_{0.2}$ ($T_0=4.6$ K, right axis) and
YbRh$_2$(Si$_{0.95}$Ge$_{0.05}$)$_2$ \cite{Kuechler} ($T_0=23.3$
K, left axis). Parallel solid lines indicate logarithmic
temperature dependence of $\beta(T)/T$.} \label{fig5}
\end{figure}


\begin{references}
\bibitem{2D Grueneisen} For 2D critical spin fluctuations an even
stronger divergence $\Gamma\propto\log\log T/T\log T$ is
predicted \cite{Zhu}.
\end{references}
\end{document}